# Laser-Assisted Metalorganic Chemical Vapor Deposition of GaN

*Yuxuan Zhang, Zhaoying Chen, Kaitian Zhang, Zixuan Feng, and Hongping Zhao[*]*

Y. Zhang, Dr. Z. Chen, K. Zhang, Z. Feng, Prof. H. Zhao
Department of Electrical and Computer Engineering, The Ohio State University, Columbus, OH 43210, USA
E-mail: zhao.2592@osu.edu

Prof. H. Zhao
Department of Materials Science and Engineering, The Ohio State University, Columbus, OH 43210, USA

[±]Y. Zhang and Dr. Z. Chen contributed equally to this work.

**Abstract:** Ammonia ($NH_3$) is commonly used as group V precursor in gallium nitride (GaN) metalorganic chemical vapor deposition (MOCVD). The high background carbon (C) impurity in MOCVD GaN is related to the low pyrolysis efficiency of $NH_3$, which represents one of the fundamental challenges hindering the development of high purity thick GaN for vertical high power device applications. This work uses a laser-assisted MOCVD (LA-MOCVD) growth technique to address the high-C issue in MOCVD GaN. Carbon dioxide ($CO_2$) laser with wavelength of 9.219 µm was utilized to facilitate $NH_3$ decomposition via resonant vibrational excitation. The LA-MOCVD GaN growth rate (as high as 10 µm/hr) shows a strong linear relationship with the trimethylgallium (TMGa) flow rate, indicating high effective V/III ratios and hence efficient $NH_3$ decomposition. Pits-free surface morphology of LA-MOCVD GaN was demonstrated for films with growth rate as high as 8.5 µm/hr. The background [C] in LA-MOCVD GaN films decreases monotonically as the laser power increases. A low [C] at $5.5\times10^{15}$ $cm^{-3}$ was achieved in LA-MOCVD GaN film grown with the growth rate of 4 µm/hr. Charge transport characterization of LA-MOCVD GaN films reveals high crystalline quality with room temperature mobility >1000 $cm^2$/Vs. LA-MOCVD growth technique provides an enabling route to achieve high quality GaN epitaxy with low-C impurity and fast growth rate simultaneously. This technique can also be extended for epitaxy of other nitride-based semiconductors.

# 1. Introduction

Gallium nitride (GaN) has played an important role in optoelectronics, photonics as well as radio-frequency (RF) electronic devices due to its wide direct band gap (3.4 eV), high electron mobility (>1000 cm$^2$/Vs), high critical electric field (3.4 MV/cm), and high thermal stability. As a promsing candidate for power electronic applications, the Baliga's figure of merit (BFOM) of GaN is more than 500X and 3X higher than that of silicon (Si) and silicon carbide (SiC). With the availability of bulk GaN substrates, vertical GaN PN diodes with breakdown voltage ($V_{BR}$) of 5 kV have been demonstrated [1]. To achieve vertical power devices with higher breakdown voltage, fundamental challenegs in GaN epitaxy with low controllable doping, thick epilayer, and high mobility still need to be addressed. GaN thin films grown by metalorganic chemical vapor deposition (MOCVD) typically have a growth rate (GR) of 2-3 µm/hr with $N_d$-$N_a$ at low-$10^{16}$ to mid-$10^{15}$ cm$^{-3}$ [2, 3]. Therefore, high quality GaN epifilm that can enable vertical GaN power devices with $V_{BR}$>10 kV is still lacking.

Trimethylgallium (TMGa) and ammonia (NH$_3$) are commonly used as the group III and group V precursors for MOCVD GaN growth. At the typical MOCVD growth temperature of GaN (~1000 °C), the pyrolysis efficiency of NH$_3$ is only at ~1% or less [4, 5, 6]. High thermal stability of NH$_3$ leads to limited N radicals for surface reaction. Therefore, the input V/III ratio during MOCVD GaN epitaxy is normally above 1000 to ensure sufficient N species reaching on the growth surface.

It has been suggested that carbon (C) impurity originating from metal-organic (MO) precursors is incorporated preferentially on nitrogen sites ($C_N$) in n-type GaN, giving rise to a deep acceptor level, which is responsible for compensating the n-type doping [7, 8]. Carbon compensation in n-type GaN drift layer based power devices can cause high leakage current and high on-resistance [9], impeding device performance. Thus, suppression C incorporation in MOCVD n-GaN epitaxy is a key research goal for high power device applications.

Carbon incorporation in MOCVD growth of III-nitrides is significantly affected by the stoichiometric balance which relies on the effective V/III ratio on the growth surface. The input V/III ratio is determined by the molar flow of group V and group III precursors, while the effective V/III ratio refers to the active group V and group III species participating during the growth process. Conventional approaches to suppress C incorporation in MOCVD GaN growth include: 1) increasing input V/III ratio via increasing NH$_3$ flow or decreasing TMGa flow [2, 9, 10, 11, 12, 13]; 2) increasing growth pressure [9, 10, 13]; and 3) increasing growth temperature [10, 13]. However, these approaches often lead to GaN growth with lower growth rates, increased gas-phase reaction, and increased thermal decomposition. Fundamentally, the low pyrolysis efficiency of $NH_3$ in MOCVD GaN growth represents a key factor that leads to relatively high-C incorporation.

Laser CVD (LCVD) is a method that introduces laser as an additional energy source during CVD process. There are two types of LCVD: 1) pyrolytic, in which a laser beam interacts with the substrate to create hot spot locally or globally for thermal-assisted chemical reactions; and 2) photolytic, in which a laser beam interacts with the chemical reactants to break the chemical bonds and thus facilitate the growth [14]. LCVD has been utilized to deposit various compound semiconductors including selective-area deposition of III-V films [15] and low-temperature MOCVD growth of GaN [16] and AlN [17]. Previously, carbon dioxide ($CO_2$) laser was also utilized in GaN MOCVD growth [18, 19, 20]. It demonstrated GaN growth on (0001) sapphire substrate at a very low temperature of 250 °C in a custom-designed system. Fast GaN growth rates of 12 µm/hr [19] and 84 µm/hr [20] were demonstrated using $CO_2$ laser at the growth temperature of 600 and 750°C, respectively. However, characterization of these materials indicates relatively low crystalline quality [19, 20].

In this work, high-quality GaN thin films using $CO_2$ laser-assisted MOCVD (LA-MOCVD) growth technique was demonstrated. The effects of the laser-assisted $NH_3$ decomposition on the GaN growth, impurity incorporation, and electron transport properties of LA-MOCVD GaN were systematically studied. Experimental results indicate that the effective V/III ratio at the growth surface can be significantly enhanced via $CO_2$ laser excitation in LA-MOCVD. C impurity incorporation can be reduced by more than 60% in LA-MOCVD GaN. GaN epifilm with low C ($5.5\times10^{15}$ cm$^{-3}$) was achieved with a low input V/III ratio of 875 and a growth rate of 4 µm/hr. LA-MOCVD GaN films show superior transport properties with room temperature mobilities over 1000 cm$^2$/Vs for films grown on both GaN-on-sapphire templates and free-standing GaN substrates with the growth rate of 4.5µm/hr. LA-MOCVD growth technique provides a new route to achieve high-quality GaN epifilms with low impurities and fast growth rates promising for high power device applications.

## 2. Results and Discussions

The schematic of the LA-MOCVD system is shown in **Figure 1**. A PRC 1500-watt tunable $CO_2$ laser was used as the laser source. A $CO_2$ laser beam is generated and introduced into the vertical rotating-disk MOCVD reactor (Agnitron Technology Inc, Agilis MOCVD

R&D System). The laser beam is parallel to the wafer surface and susceptor. A beam dump is placed at the front viewport to absorb the residual laser power.

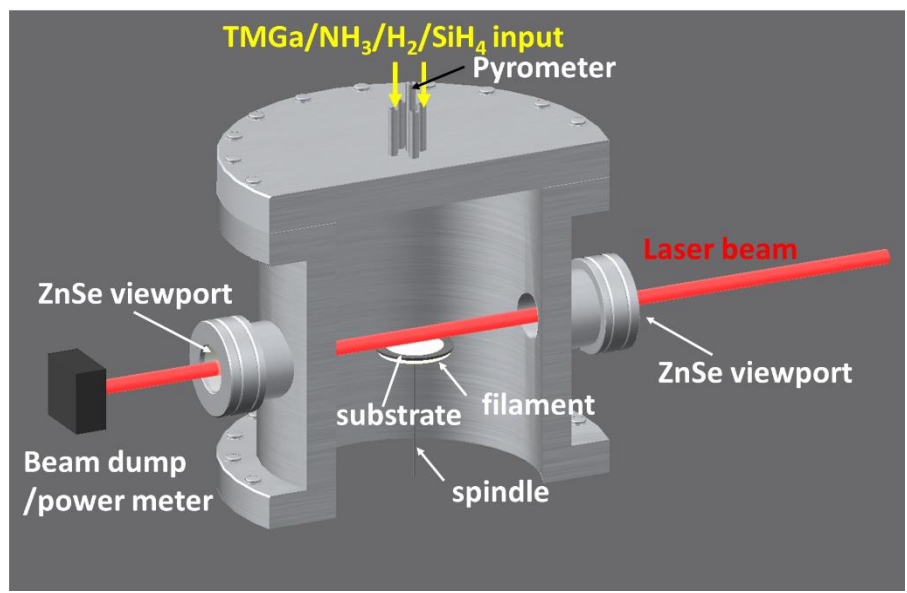

**Figure 1.** Schematic illustration of LA-MOCVD system.

**2.1 Interaction between $CO_2$ laser and $NH_3$**

$CO_2$ laser power absorption under different controlled chamber conditions was investigated. **Figure 2**(a) shows the laser power absorption as a function of the laser wavelength under a typical reactor condition for GaN growth (965 °C, 200 torr, 4 standard liter per minute (slm) $NH_3$ flow and 4 slm $H_2$ flow). The input laser power was fixed at 200W. The most prominent absorption was observed at 9.219 µm (1084.71cm$^{-1}$) where ~100% of the input laser power was absorbed. This wavelength matches with the rotational−vibrational transition of the N-H wagging mode ($v_2$) of the $NH_3$ molecules at 1084.63 cm$^{-1}$ [19, 20, 21]. Therefore, 9.219 µm laser wavelength was selected as the operating wavelength in the following investigations. Figure 2(b) plots the laser power absorption (%) as a function of the $NH_3$ flow rate under chamber pressure of 200 Torr, with a fixed laser wavelength at 9.219 µm and laser power of 200 W. The laser power absorption increases as the $NH_3$ flow rate increases with a quick saturation at $NH_3$ flow rate of 0.75 slm. Further increase of $NH_3$ flow rate leads to a full absorption of the incident laser power. Our study also reveals a strong dependence of laser power absorption on the chamber pressure. As shown in Figure 2(c), with a fixed $NH_3$ flow rate of 4 slm, the laser absorption was measured as a function of input laser power under different chamber pressure at 75, 100, 150 and 200 Torr. For the case of 75 Torr, the laser beam was fully absorbed when the input laser power was below 150 W. As the input laser power further increases beyond this threshold, the absorption % shows a gradual decrease. The similar trend was observed for all cases with different chamber pressure, but the input laser power threshold increases as the chamber pressure increases. With a chamber pressure of 200 Torr, the input laser power threshold was measured at 350 W. From Figure 2(d), the threshold input laser power shows a linear relationship with the chamber pressure. According to the ideal gas law: $PV=nRT$, with fixed temperature ($T$), reactor volume ($V$), and gas composition, the amount of $NH_3$ molecules ($n$) is proportional to the chamber pressure ($P$). Therefore, the threshold input power is proportional to the amount of $NH_3$ molecules along the beam pathway.

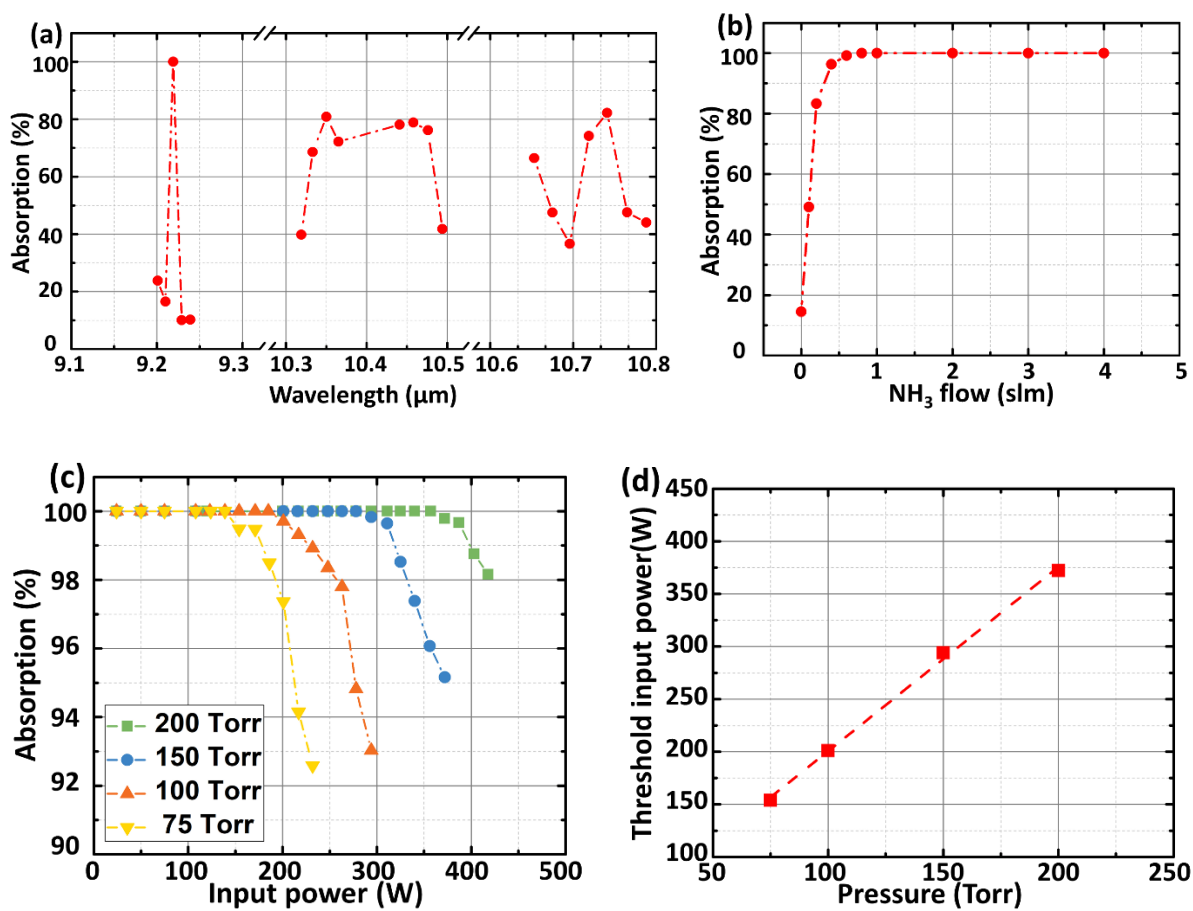

**Figure 2.** (a) Absorption spectrum of $CO_2$ laser power through MOCVD reactor with 4 slm $NH_3$ flow rate and 150 torr chamber pressure. (b) Laser power absorption as a function of $NH_3$ flow rate with 200 W input power and 9.219 µm wavelength. The chamber pressure was kept at 150 torr. (c) Laser power absorption as a function of input laser power under different chamber pressures with 200 W input power and 9.219 µm wavelength. The $NH_3$ flow rate was kept at 4slm. (d) The penetration threshold of input laser power as a function of chamber pressure extracted from (c).

From the above studies, $CO_2$ laser beam at wavelength of 9.219 µm has a strong interaction with $NH_3$ molecules. The laser power absorption is strongly dependent on $NH_3$ flow rate and chamber pressure. The following sections discuss the GaN growth via LA-MOCVD method and the comprehensive material characterization.

**2.2 Growth of LA-MOCVD GaN**

Typical MOCVD GaN growth is performed under N-rich conditions with the input V/III ratio up to several thousand. Thus, the GaN growth rate is limited by the mass transport of Ga species, i.e., the TMGa flow, and therefore has a linear relationship with the TMGa flow rate [12]. As TMGa flow rate increases, the effective V/III ratio decreases, which can lead to non-linear relationship between the growth rate and TMGa flow rate due to the limited N species. **Figure 3** shows the GaN growth rate as a function of the TMGa flow rate under different input laser power: 0 W (without laser), 150 W and 250 W. The chamber pressure was fixed at 150 Torr and $NH_3$ flow was kept at 4 slm. For the case without laser excitation (0 W), GaN growth rate shows a trend of saturation as the TMGa flow rate increases (>0.4 mmol min$^{-1}$), indicating the reduction of the effective V/III ratio. For the cases with laser excitation (150 W and 250 W), the GaN growth rate increases linearly as the TMGa flow rate increases, indicating the GaN growths still fall under N-rich conditions. Thus, the introduction of laser beam during GaN MOCVD growth can effectively enhance $NH_3$ cracking efficiency. Note that the slight reduction of the LA-MOCVD GaN growth rates as compared to those of the conventional GaN growth (0 W) is due to the increased gas phase reaction during LA-MOCVD process. Nevertheless, LA-MOCVD growth technique is a viable approach to achieve fast growth rate MOCVD GaN with relatively high effective V/III ratio.

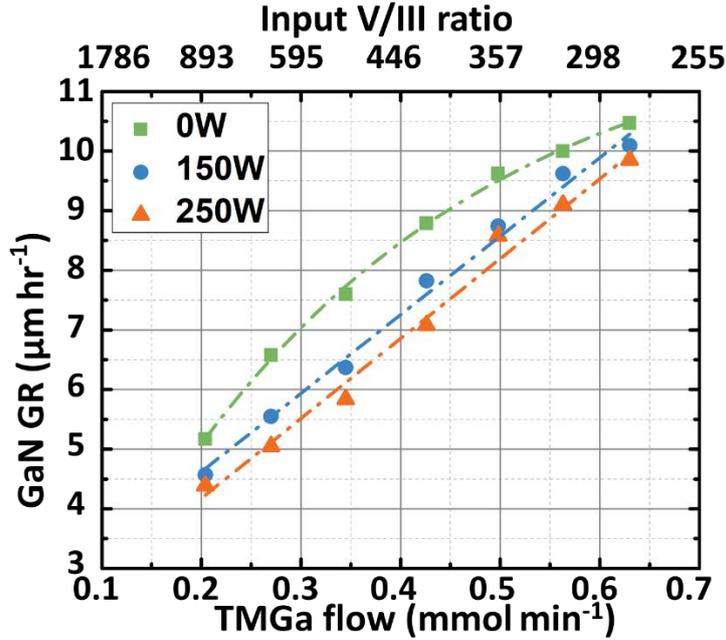

**Figure 3.** GaN growth rate as a function of TMGa flow rate with 0 W, 150 W and 250 W input laser power. The NH₃ flow rate was fixed at 4 slm and chamber pressure at 150 torr.

Surface morphologies of two GaN samples grown without and with laser excitation (250 W) were compared. The MOCVD growth condition including chamber pressure (150 Torr), input V/III ratio (415), and growth temperature (1025 °C) was kept the same for both samples. The corresponding GaN growth rates were 9.3 μm hr$^{-1}$ and 8.5 μm hr$^{-1}$ for the processes without and with laser, respectively. The total GaN film thickness was ~ 2.8 μm for both samples. **Figures 4**(a) and 4(b) compare the macroscopic surface morphology of these two samples via optical imaging. Similar stripe morphologies were observed on both samples, indicating minimum impact of the laser excitation on the macroscopic surface morphology. Figures 4(c) and 4(d) compare the SEM images taken from the same two samples. Hexagonal shaped V-pits were observed on the GaN sample grown without the laser (Figure 4(c)). V-pits are commonly observed in N-deficient MOCVD GaN growth, indicating a low effective V/III ratio during the epitaxy. In contrast, the LA-MOCVD grown GaN sample shows a smooth surface without observable V-pits, indicating a high effective V/III ratio due to efficient NH₃ cracking from the laser excitation. Figures 4(e) and 4(f) show the comparison of the 10×10 μm² AFM imaging between the two GaN films. Both films were grown under step growth mode with low RMS values of 0.606 nm (without laser) and 0.581 nm (with laser). From the investigation of the GaN surface morphologies, it is further confirmed that LA-MOCVD provides more efficient NH₃ cracking with the laser excitation, which facilitates GaN growth with higher V/III ratio.

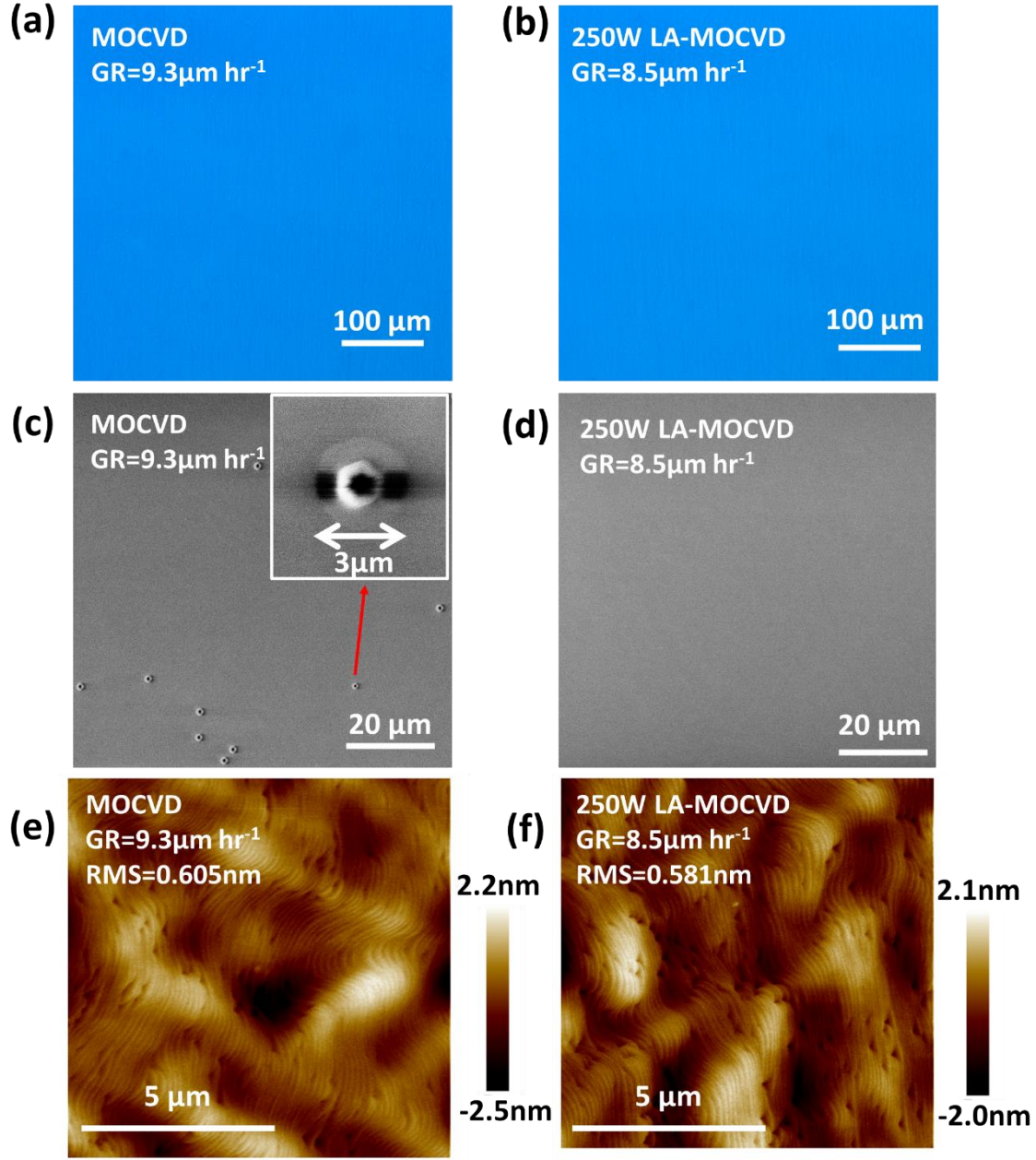

**Figure 4.** (a, b) Optical microscope, (c, d) SEM and (e, f) 10×10 µm$^2$ AFM images of surface morphology of GaN sample grown by conventional MOCVD (a, c, e) and LA-MOCVD (b, d, f). The MOCVD growth condition including chamber pressure (150 Torr), input V/III ratio (415), and growth temperature (1025 °C) was kept the same for both samples. The growth rate of MOCVD and LA-MOCVD sample were 9.3 and 8.5 µm hr$^{-1}$, respectively.

## 2.3 Impurity incorporation

Analytical models based on the Langmuir isotherm have been utilized to study the surface reconstructions and carbon coverages in GaN MOCVD under different growth conditions [22, 23]. It was found that the surface carbon coverage is positively related to the carbon incorporation in GaN. In this work, a theoretical Langmuir adsorption model is utilized to investigate the effect of laser-assisted $NH_3$ decomposition on the GaN growth surface carbon coverage.

At equilibrium, the fractional coverage $\theta$ of an adsorbate can be expressed as [24]:

$$\theta = \frac{\exp(\frac{\mu_{gas}+\varepsilon}{kT})}{1+\exp(\frac{\mu_{gas}+\varepsilon}{kT})} \qquad (1)$$

where $\mu_{gas}$ is the chemical potential of gas species, $\varepsilon$ is the absorption energy. For a multi-species absorption process, the fractional coverage of the $i^{th}$ adsorbate can be re-written as:

$$\theta_i = \frac{\sum_i \frac{p_i}{p^0}\exp(\frac{\mu^0_{gas,i}+\varepsilon_i}{kT})}{1+\sum_j \frac{p_j}{p^0}\exp(\frac{\mu^0_{gas,j}+\varepsilon_j}{kT})} \qquad (2)$$

where $p_i$ represents the partial pressure of the $i^{th}$ gas species and $p^0$ represents the reference (atmosphere) pressure. To qualitatively illustrate how the laser-assisted $NH_3$ decomposition influences the carbon incorporation, key absorption processes were taken into consideration in this model. The absorption processes and the corresponding absorption energies cited from prior density-functional theory (DFT) reports are listed in Table 1. Three N-related processes (the absorption of $NH_3$, $NH_2$, and NH) and three Ga-related processes (Ga absorption on different surface sites) are considered. Regarding C absorption, $CH_3Ga$ (MMGa) and $CH_3$ were suggested as the main sources [25, 26], which were taken into account in this model. It is worth noting that the simplified model in this work intends to qualitatively study the impact of the $NH_3$ decomposition on C incorporation in GaN MOCVD.

Assuming the $NH_3$ radicals ($NH_2$ and NH) are from the direct decomposition of $NH_3$, a two-step decomposition processes are considered: 1) First-step decomposition: $NH_3(g) \leftrightarrow NH_2(g) + H(g)$; and 2) Second-step decomposition: $NH_2(g) \leftrightarrow NH(g) + H(g)$. Figure 5 plots the carbon coverage as a function of $NH_3$ decomposition percentage. Typical MOCVD GaN growth condition (growth temperature of 1050°C, growth pressure of 200 torr, and input V/III ratio of 1000) was used in this model. The chemical potentials of gas-phase species were taken from Ref. [27]. Here, $a_0$ and $a_1$ represent the equilibrium conversion percentage of each reaction. For example, $a_1$=1% and $a_2$=1% means 1% of $NH_3$ decomposes into $NH_2$, and 1% of the produced $NH_2$ decomposes into NH. **Figure 5** shows that surface C coverage can be greatly influenced by $NH_3$ decomposition. As $a_1$ and $a_2$ increase from 1% (typical for $NH_3$ decomposition at ~1000 °C) to 100%, the C coverage reduces by more than 3 orders of magnitude. Note that $NH_2$ and NH have higher chemical potentials than $NH_3$, suggesting they are more chemically active. Furthermore, the larger adsorption energies of $NH_2$ and NH as compared to NH$_3$ enable higher adsorption on the growth surface. Both theoretical [7] and experimental [8] studies show that C prefers to occupy N sites and form a deep acceptor state (C$_N$). Since C and N adatoms compete to occupy the same sites on the growth surface, more absorption of NH$_2$ and NH results in less C incorporation. Although complex gas-phase and surface reactions can influence the absolute C incorporation in GaN MOCVD growth process, this simplified analytical model indicates that enhancing NH$_3$ decomposition is a promising route to suppress C incorporation in GaN MOCVD.

**Table 1.** N, Ga, and C related absorption processes and their corresponding absorption energies calculated by DFT from literatures.

| Absorption process | Absorption energy (eV) | Reference |
|---|---|---|
| NH$_3$(gas)↔NH$_3$(ad, T1) | 1.17 | [28] |
| NH$_2$(gas)↔NH$_2$(ad, Br) | 3.97 | |
| NH(gas)↔NH(ad,H3) | 6.43 | |
| Ga (gas)↔ Ga (ad, T1) | 3.77 | [29] |
| Ga (gas)↔ Ga (ad, T4) | 2.48 | |
| Ga (gas)↔ Ga (ad, H3) | 3.49 | |
| MMGa (gas)↔ MMGa (ad, T1) | 2.83 | [30] |
| CH$_3$ (gas)↔ CH$_3$ (ad) | 5.07 | [25] |

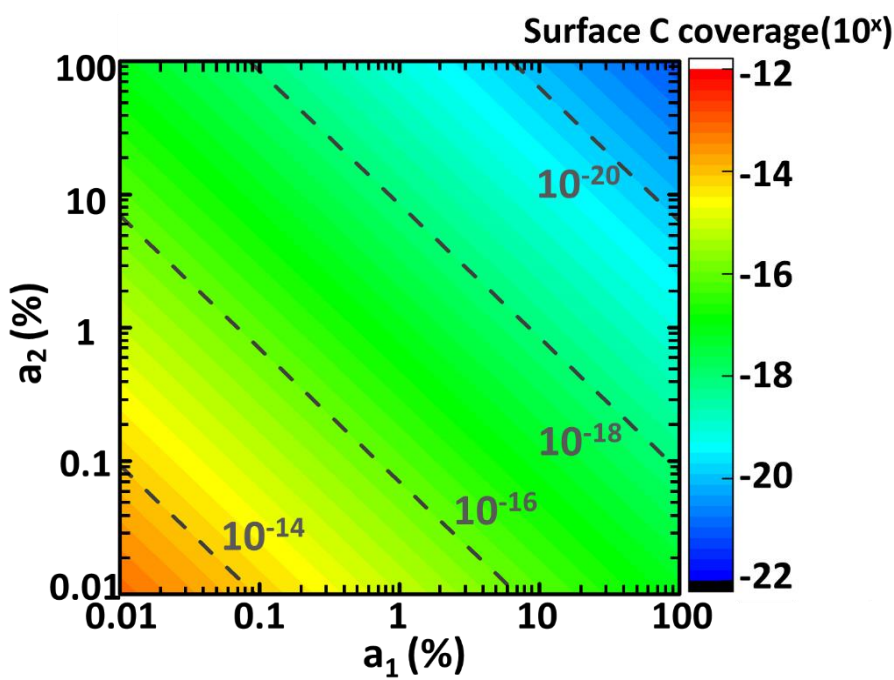

**Figure 5.** Contour of surface carbon coverage as a function of the first- and second-step of NH$_3$ decomposition obtained from theoretical modeling. 1050 °C temperature, 200 torr pressure and 1000 input V/III ratio were selected as the modeling condition.

A series of GaN sub-layers were grown under various growth conditions for quantitative SIMS characterization to probe C impurity concentration. The detailed growth conditions were listed in Table 2, which were categorized in two groups: Group I (growth pressure 150 Torr) and Group II (growth pressure 200 Torr). Among each group, the input laser power was varied to investigate the impact of laser power on C incorporation and GaN growth rate. **Figure 6** plots the C concentration and GaN growth rate as a function of the laser power. Cases from Group I show that [C] in GaN layers decreases as the laser power increases. [C] was reduced from $2.5\times10^{16}$ cm$^{-3}$ (0 W) to $1.2\times10^{16}$ cm$^{-3}$ (250 W). The GaN growth rate among Group I reduced from 5.2 µm hr$^{-1}$ (0 W) to 4.6 µm hr$^{-1}$ (150 W) and 4.4 µm hr$^{-1}$ (250 W). The slight reduction in growth rate is due to enhanced gas phase reaction as discussed above. A more prominent [C] reduction in LA-MOCVD GaN growth was observed from cases in Group II, in which the growth pressure was set as 200 Torr. The [C] was reduced from $1.4\times10^{16}$ cm$^{-3}$ (0 W) to $5.5\times10^{15}$ cm$^{-3}$ (200 W), whereas the growth rate decreased from 5.2 µm hr$^{-1}$ (0 W) to 4 µm hr$^{-1}$ (200 W). This group of data directly demonstrate the effective reduction of [C] in LA-MOCVD GaN growth due to enhanced NH$_3$ cracking efficiency.

**Table 2**. Growth conditions of GaN SIMS stacks with the corresponding growth rates and carbon concentrations.

| Group | Pressure (torr) | V/III ratio | Laser power (W) | Growth rate (µm hr$^{-1}$) | [C] ($\times10^{16}$ cm$^{-3}$) |
|---|---|---|---|---|---|
| Group I | 150 | 876 | 0 | 5.2 | 2.5 |
| | | | 50 | 5.2 | 2.4 |
| | | | 150 | 4.6 | 1.5 |
| | | | 250 | 4.4 | 1.2 |
| Group II | 200 | | 0 | 5.2 | 1.5 |
| | | | 200 | 4 | 0.55 |

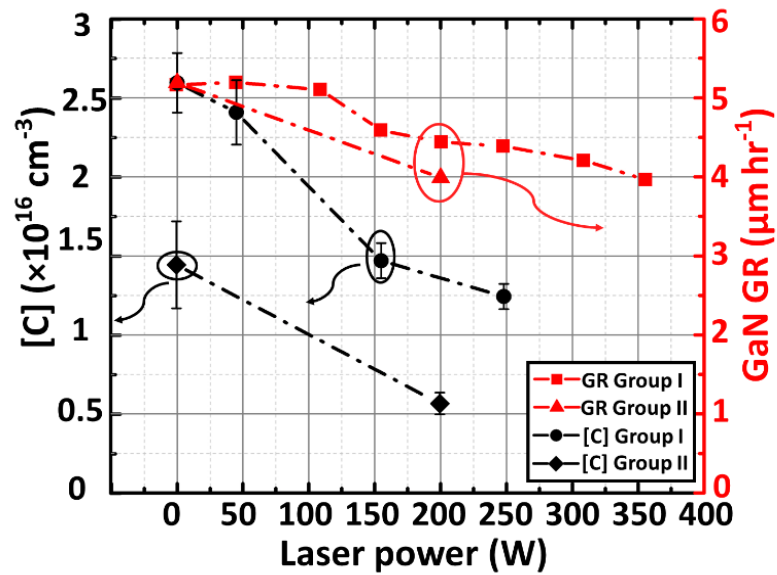

**Figure 6.** [C] and growth rates of LA-MOCVD GaN as a function of input laser power. The growth conditions of the samples are listed in Table 2.

**Figure 7**(a) summarizes the [C] in LA-MOCVD GaN from this work as compared to the representative [C] in reported MOCVD GaN growth as a function of the input V/III ratio. The dashed line indicates the slope of [C] vs. input V/III ratio among the best reported data, considering [C] is reversely proportional to the input V/III ratio. One can observe that significantly high V/III ratios are required to achieve [C] < $10^{16}$ cm$^{-3}$. Using LA-MOCVD growth technique, the achieved [C] at ~$5.5\times10^{15}$ cm$^{-3}$ with a V/III ratio at 875 represents the lowest value as compared to those grown at the similar V/III ratios.

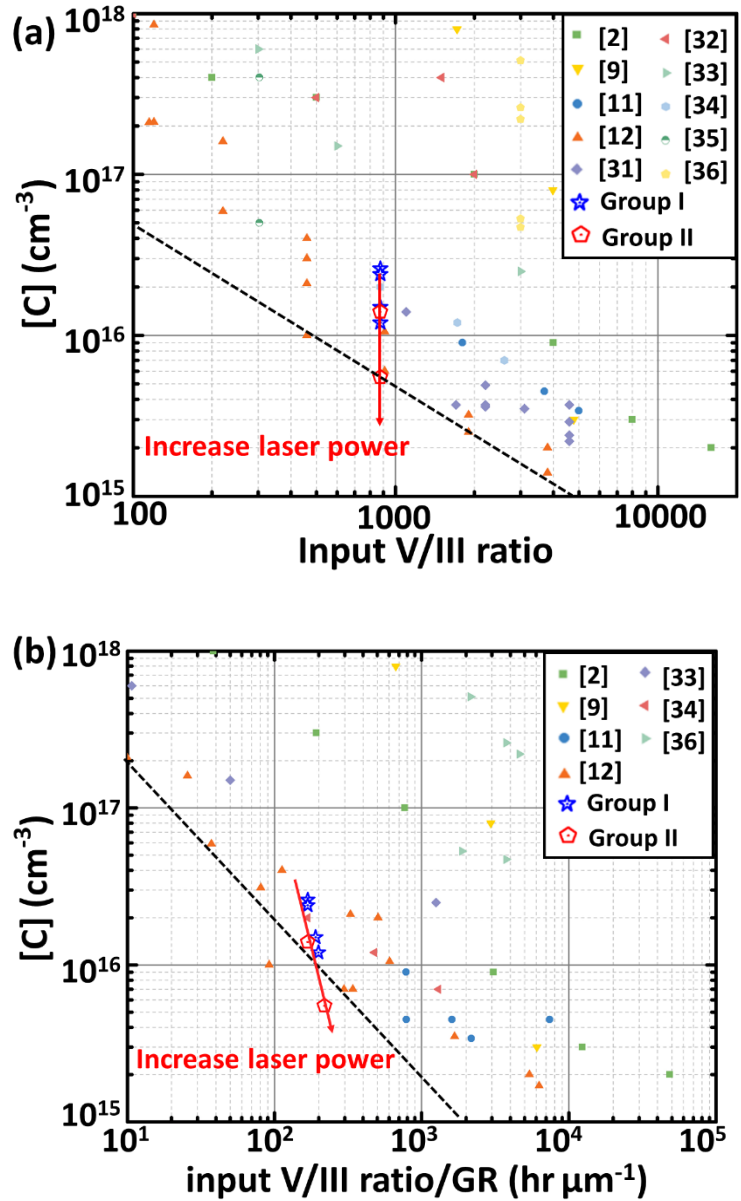

**Figure 7.** (a) [C] versus input V/III ratio of MOCVD GaN from literatures [2, 9, 11, 12, 31, 32, 33, 34, 35, 36] and from this study. (b) [C] versus input V/III ratio/growth rate of MOCVD GaN from literatures [2, 9, 11, 12, 33, 34, 36] and from this study. The growth conditions, corresponding growth rate and [C] of the LA-MOCVD samples are listed in Table 2.

In addition to the V/III ratio, the absolute flow rate of TMGa or $NH_3$ can significantly impact the background [C] incorporation as well. This factor can be taken into account from the GaN film growth rate, considering [C] is positively related to the GaN growth rate [12, 34]. Therefore, [C] is plotted as a function of the input V/III ratio/GR from this work as compared to data reported from literature, as shown in Figure 7(b). The dashed line indicates the slope of [C] vs. input V/III ratio/GR for the best reported data, considering [C] is reversely proportional to V/III ratio and 1/GR. It illustrates the trade-off between [C] and growth rate as well as the V/III ratio. In conventional MOCVD GaN growth, in order to achieve low [C] impurity, low growth rate and high V/III ratio is necessary. In contrast, LA-MOCVD GaN growth enables low [C] with fast growth rate. With the increase of laser power, LA-MOCVD showed significant [C] reduction without notable increase of input V/III ratio/GR. This is due to the high effective V/III ratio at the growth surface through laser-assisted $NH_3$ decomposition. As compared to previous reports, LA-MOCVD GaN growth achieved low [C] at mid-$10^{15}$ cm$^{-3}$ with the lowest input V/III ratio/GR.

## 2.4 Transport properties

**Figure 8**(a) compares the room temperature mobilities of MOCVD grown GaN films on both GaN/sapphire templates and free-standing GaN substrates, with and without laser excitation. The growth rates of conventional MOCVD and LA-MOCVD GaN films were 5.2 $\mu m\,hr^{-1}$ and 4.5 $\mu m\,hr^{-1}$, respectively. The trend indicates that GaN films grown via LA-MOCVD have superior room temperature mobilities as compared to those grown without laser excitation. The electron mobilities of GaN films grown without laser excitation ranged from 600 to 750 cm$^2$ V$^{-1}$ s$^{-1}$. LA-MOCVD GaN films show higher electron mobilities ranging between 750 and 1275 cm$^2$V$^{-1}$s$^{-1}$. The enhanced electron mobilities from LA-MOCVD GaN samples are most likely due to the reduced impurity scattering effect from lower carbon concentration [37]. Figure 8(b) compares the room temperature electron mobilities of GaN grown via LA-MOCVD from this work as compared to those grown via molecular beam epitaxy (MBE), halide vapor phase epitaxy (HVPE), and regular MOCVD. The LA-MOCVD GaN films are among the few reports that achieved room temperature mobility over 1000 cm$^2$ V$^{-1}$ s$^{-1}$. More importantly, LA-MOCVD GaN films contain a growth rate at 4.5 $\mu m\,hr^{-1}$, much higher than those grown via MBE or conventional MOCVD conditions.

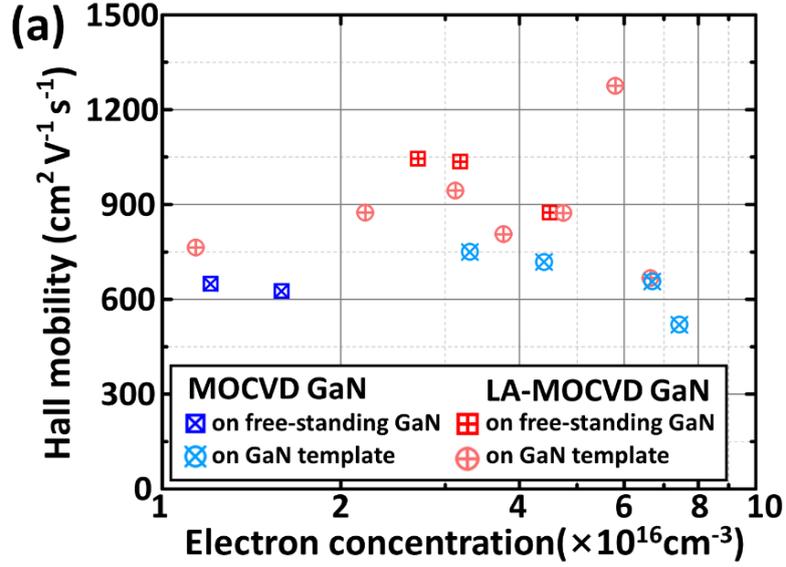

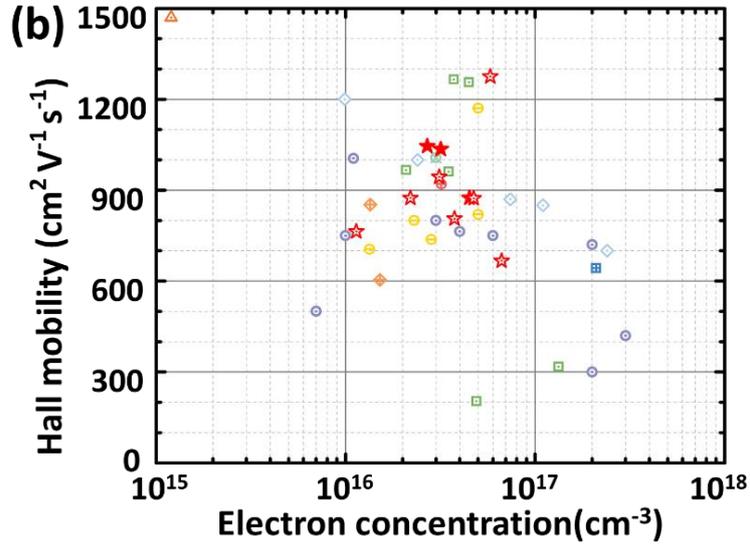

| Epitaxy method | Substrate | GR (μm hr⁻¹) | Legend | reference |
|---|---|---|---|---|
| NH₃-based MBE | GaN template | 0.4 | ▫ | [38] |
| MBE | Sapphire | Not reported | ⊞ | [39] |
| Quartz-free HVPE | Free-standing GaN | Not reported | △ | [40] |
| MOCVD | Sapphire/ GaN template | 0.33 - 1.3 | ⊖ | [2] |
| | | Not reported | ⊙ | [41] |
| | | Not reported | ⊕ | [42] |
| | | 2 | ⊠ | [3] |
| | Free-standing GaN | Not reported | ◇ | [31] |
| | | 2 - 5.2 | ⊕ | [34] |
| LA-MOCVD | Free-standing GaN | 4.5 | ★ | This work |
| | GaN template | 4.5 | ☆ | |

**Figure 8.** (a) Hall mobility of GaN samples grown with conventional MOCVD and LA-MOCVD with electron concentration between $10^{16}$ to $10^{17}$ cm$^{-3}$. The growth rates of conventional MOCVD and LA-MOCVD samples were 5.2 and 4.5 μm hr$^{-1}$, respectively. (b) Hall mobility versus electron concentration of GaN grown with HVPE, MBE and MOCVD on various substrates from literatures [2, 3, 31, 34, 38, 39, 40, 41, 42], and LA-MOCVD samples from this study. The reported growth rates from literatures are listed in the table.

## 3. Conclusion

In summary, LA-MOCVD GaN epitaxy was developed to facilitate $NH_3$ decomposition and thus to achieve high quality GaN films with low-C and fast growth rate. It is confirmed that $NH_3$ molecules are most efficiently decomposed with $CO_2$ laser excitation at 9.219 µm. The strong linear relationship of the GaN growth rate (as high as 10 µm hr$^{-1}$) with TMGa flow rate and pits-free surface morphology at high growth rate (8.3 µm hr$^{-1}$) from LA-MOCVD GaN indicate high effective V/III ratios due to efficient $NH_3$ decomposition. The background [C] in LA-MOCVD GaN films decreased monotonically with the increase of the laser power. Low [C] at $5.5\times10^{15}$ cm$^{-3}$ was achieved in LA-MOCVD GaN film grown with the growth rate of 4 µm/hr. LA-MOCVD GaN exhibited lower [C] incorporation at low input V/III ratio and high growth rates as compared to the conventional MOCVD GaN growth. High room temperature mobilities over 1000 cm$^2$ V$^{-1}$ s$^{-1}$ were achieved by LA-MOCVD with a growth rate of 4.5 µm hr$^{-1}$ grown on both free-standing GaN substrate and GaN/sapphire template. LA-MOCVD GaN growth technique provides a promising approach to obtain high-quality, low-C GaN with fast growth rate for power device applications. The similar method can also be extended to the epitaxy of other nitride semiconductors.

## 4. Experimental section

GaN films were grown on 2" (0001) GaN-on-sapphire templates (GaN templates) and ammonothermal Mn-doped free-standing GaN substrates in a commercial vertical rotating-disk MOCVD reactor (Agnitron Technology Inc, Agilis MOCVD R&D System). The temperature of growth surface was measured in-situ via a pyrometer. TMGa and $NH_3$ were used as Ga and N precursors, respectively. $H_2$ was used as the carrier gas. Diluted silane ($SiH_4$) balanced in $N_2$ was used as Si dopant. Impurities of selected samples were characterized by quantitative secondary ion mass spectroscopy (SIMS), with the carbon detection limit (DL) of $1\times10^{15}$ cm$^{-3}$. Surface morphologies of GaN films were characterized by Huvitz HRM-300 optical microscope, Thermo Scientific Apreo scanning electron microscopy (SEM), and Bruker atomic force microscopy (AFM). Room temperature electron transport properties were measured by Van-der-Pauw Hall configuration using the Ecopia HMS-3000 Hall effect system.

**Acknowledgments**

The information, data, or work presented herein was funded in part by the Advanced Research Projects Agency-Energy (ARPA-E), U.S. Department of Energy, under Award Number DE-AR0001036, and U.S. Department of Energy's Office of Energy Efficiency and Renewable Energy (EERE) under the Advanced Manufacturing Office, FY18/FY19 Lab Call. The views and opinions of authors expressed herein do not necessarily state or reflect those of the United States Government or any agency thereof.

Received: ((will be filled in by the editorial staff))
Revised: ((will be filled in by the editorial staff))
Published online: ((will be filled in by the editorial staff))

**Data Availability**

The data that support the findings of this study are available from the corresponding author upon reasonable request.

**References**

[1] H. Ohta, K. Hayashi, F. Horikiri, M. Yoshino, T. Nakamura, and T. Mishima, *Jpn. J. Appl. Phys.* **2018**, *57*, 04FG09.

[2] F. Kaess, S. Mita, J. Xie, P. Reddy, A. Klump, L. H. Hernandez-Balderrama, S. Washiyama, A. Franke, R. Kirste, A. Hoffmann, and R. Collazo, *J. Appl. Phys.* **2016**, *120*, 105701.

[3] Y. Zhang, Z. Chen, W. Li, H. Lee, M. R. Karim, A. R. Arehart, S. A. Ringel, S. Rajan, and H. Zhao, *J. Appl. Phys.* **2020**, *127*, 215707.

[4] A. H. White, and W. Melville, *J. Am. Chem. Soc.* **1905**, *27*, 373.

[5] V. S. Ban, *J. Electrochem. Soc.* **1972**, *119*, 761.

[6] D. F. Davidson, K. Kohse-Höinghaus, A. Y. Chang, and R. K. Hanson, *Int. J. Chem. Kinet.* **1990**, *22*, 513.

[7] J. L. Lyons, A. Janotti, and C. G. Van de Walle, *Phys. Rev. B* **2014**, *89*, 035204.

[8] A. Armstrong, A. R. Arehart, B. Moran, S. P. DenBaars, U.K. Mishra, J. S. Speck, and S. A. Ringel, *Appl. Phys. Lett.* **2004**, *84*, 374.

[9] Y. Cao, R. Chu, R. Li, M. Chen, R. Chang, and B. Hughes, *Appl. Phys. Lett.* **2016**, *108*, 062103.

[10] D. D. Koleske, A. E. Wickenden, R. L. Henry, and M. E. Twigg, *J. Cryst. Growth* **2002**, *242*, 55.

[11] G. Piao, K. Ikenaga, Y. Yano, H. Tokunaga, A. Mishima, Y. Ban, T. Tabuchi, and K. Matsumoto, *J. Cryst. Growth* **2016**, *456,* 137.

[12] T. Ciarkowski, N. Allen, E. Carlson, R. McCarthy, C. Youtsey, J. Wang, P. Fay, J. Xie, and L. Guido, *Materials* **2019**, *12*, 2455.

[13] N. A. Fichtenbaum, T. E. Mates, S. Keller, S. P. DenBaars, and U. K. Mishra, *J. Cryst. Growth* **2008**, *310*, 1124.

[14] J. Mazumder, and A. Kar, *Theory and application of laser chemical vapor deposition*, Springer Science & Business Media, New York, NY **2013**.

[15] R. Iga, H. Sugiura, and T. Yamada, *Semicond. Sci. Technol.* **1993**, *8*, 1101.

[16] B. Zhou, Z. Li, T. L. Tansley, and K. S. A. Butcher, *J. Cryst. Growth* **1996**, *160*, 201.

[17] X. Li, and T. L. Tansley, *J. Appl. Phys.* **1990**, *68*, 5369.

[18] H. Rabiee Golgir, D. W. Li, K. Keramatnejad, Q. M. Zou, J. Xiao, F. Wang, L. Jiang, J. F. Silvain, and Y. F. Lu, *ACS Appl. Mater. Interfaces* **2017**, *9*, 21539.

[19] H. Rabiee Golgir, Y. Gao, Y. S. Zhou, L. Fan, P. Thirugnanam, K. Keramatnejad, L. Jiang, J. F. Silvain, and Y. F. Lu, *Cryst. Growth Des.* **2014**, *14*, 6248.

[20] H. Rabiee Golgir, Y. S. Zhou, D. Li, K. Keramatnejad, W. Xiong, M. Wang, L. J. Jiang, X. Huang, L. Jiang, J. F. Silvain, and Y. F. Lu, *J. Appl. Phys.* **2016**, *120*, 105303.

[21] J. McBride and R. Nicholls, *J. Phys. B* **1972**, *5*, 408.

[22] Y. Inatomi, and Y. Kangawa, *Appl. Surf. Sci.* **2020**, *502*, 144205.

[23] D. Yosho, Y. Inatomi, and Y. Kangawa, *Jpn. J. Appl. Phys.* **2020**, *59*, 048002.

[24] K. W. Kolasinski, *Surface science: foundations of catalysis and nanoscience*, John Wiley & Sons, West Sussex, England, **2012**.

[25] Ö. Danielsson, X. Li, L. Ojamäe, E. Janzén, H. Pedersen, and U. Forsberg, *J. Mater. Chem. C* **2016**, *4*, 863.

[26] X. Li, Ö. Danielsson, H. Pedersen, E. Janzén, and U. Forsberg, *J. Vac. Sci. Technol. B* **2015**, *33*, 021208.

[27] I. Barin, *Thermochemical data of pure substances*, VCH, New York, NY **1989**.

[28] H. Suzuki, R. Togashi, H. Murakami, Y. Kumagai, and A. Koukitu, *Phys. Status Solidi C* **2009**, *6*, S301.

[29] K. M. Bui, J. I. Iwata, Y. Kangawa, K. Shiraishi, Y. Shigeta, and A. Oshiyama, *J. Phys. Chem. C* **2018**, *122*, 24665.

[30] Y. S. Won, J. Lee, C. S. Kim, and S. S, Park, *Surf. Sci.* **2009**, *603*, L31.

[31] N. Sawada, T. Narita, M. Kanechika, T. Uesugi, T. Kachi, M. Horita, T. Kimoto, and J. Suda, *Appl. Phys. Express* **2018**, *11*, 041001.

[32] S. Mita, R. Collazo, A. Rice, R. F. Dalmau, and Z. Sitar, *J. Appl. Phys.* **2008**, *104*, 013521.

[33] K. Matsumoto, H. Tokunaga, A. Ubukata, K. Ikenaga, Y. Fukuda, T. Tabuchi, Y. Kitamura, S. Koseki, A. Yamaguchi, and K. Uematsu, *J. Cryst. Growth* **2008**, *310*, 3950.

[34] Y. Zhang, Z. Chen, W. Li, A. R. Arehart, S. A. Ringel, and H. Zhao, *Phys. Status Solidi a*, https://doi.org/10.1002/pssa.202000469

[35] A. E. Wickenden, D. D. Koleske, R. L. Henry, M. E. Twigg, and M. Fatemi, *J. Cryst. Growth* **2004**, *260*, 54.

[36] J. Yang, D. G. Zhao, D. S. Jiang, P. Chen, Z. S. Liu, L. C. Le, X. J. Li, X. G. He, J. P. Liu, S. M. Zhang, and H. Wang, *J. Appl. Phys.* **2014**, *115*, 163704.

[37] K. Seeger, *Semiconductor physics – An Introduction*, Springer-Verlag Berlin Heidelberg, New York, NY 2013

[38] E. C. Kyle, S. W. Kaun, P. G. Burke, F. Wu, Y. R. Wu, and J. S. Speck, *J. Appl. Phys.* **2014**, *115*, 193702.

[39] D. C. Look and J. R. Sizelove, *Phys. Rev. Lett.* **1999**, *82*, 1237.

[40] H. Fujikura, T. Konno, T. Kimura, Y. Narita, and F. Horikiri, *Appl. Phys. Lett.* **2020**, *117*, 012103.

[41] D. G. Zhao, H. Yang, J. J. Zhu, D. S. Jiang, Z. S. Liu, S. M. Zhang, Y. T. Wang, and J. W. Liang, *Appl. Phys. Lett.* **2006**, *89*, 112106.

[42] S. Nakamura, T. Mukai, and M. Senoh, *J. Appl. Phys.* **1992**, *71*, 5543.